# Application of Computer Deep Learning Model in Diagnosis of Pulmonary Nodules

Yutian Yang[1], Hongjie Qiu[2], Yulu Gong[3], Xiaoyi Liu[4], Yang Lin[5], Muqing Li[6]

[1]University of California, Davis,USA ,yytyang@ucdavis.edu

[2]University of Washington,USA,hongjieq@uw.edu

[3]Northern Arizona University,USA,yg486@nau.edu

[4]Arizona State University,USA,xliu472@asu.edu

[5]University of Pennsylvania,USA,yang.lin1345@gmail.com

[6]University of California San Diego,USA,MUL003@ucsd.edu

*Abstract*—The 3D simulation model of the lung was established by using the reconstruction method. A computer aided pulmonary nodule detection model was constructed. The process iterates over the images to refine the lung nodule recognition model based on neural networks. It is integrated with 3D virtual modeling technology to improve the interactivity of the system, so as to achieve intelligent recognition of lung nodules. A 3D RCNN (Region-based Convolutional Neural Network) was utilized for feature extraction and nodule identification. The LUNA16 large sample database was used as the research dataset. FROC (Free-response Receiver Operating Characteristic) analysis was applied to evaluate the model, calculating sensitivity at various false positive rates to derive the average FROC. Compared with conventional diagnostic methods, the recognition rate was significantly improved. This technique facilitates the detection of pulmonary abnormalities at an initial phase, which holds immense value for the prompt diagnosis of lung malignancies.

## I. INTRODUCTION

In recent years, medical images represented by CT have gradually become an effective clinical examination method. Lung cancer has the highest incidence rate among all malignant tumors. Pulmonary nodules are a common lung condition, and though only a small fraction are malignant, diagnosing them is crucial due to their potentially indicative nature[1]. Currently, CT scans are the most effective means of diagnosing lung cancer. However, relying solely on manual examination by doctors is not only labor-intensive but also susceptible to errors.

Some researchers have proposed a special point enhancement filtering algorithm and constructed a neural network using a manually extracted feature set to weed out spurious nodes [2]. Deep learning has rapidly advanced in imaging, video, and other fields, with scholars beginning to apply these techniques to lung lesion identification, yielding promising results[3]. At present, it has been proposed to use two-dimensional convolutional networks to identify lung lesions, and its recognition accuracy is 15-20% higher than that of conventional methods [4]. However, due to its continuous characteristics in space, pulmonary nodule in CT image is regarded as a 3D object recognition, so the traditional convolutional neural network method cannot extract the features of lung CT image well [5]. Some researchers have proposed a 3D model training method based on convolutional networks, which can extract the spatial features of 3D objects well. Compared with 2D CNN method, 3D CNN has greatly improved the screening accuracy of false positive nodes[6]. In order to solve the above problems, this project intends to study a new algorithm of deep convolutional neural network to identify pulmonary nodules.

## II. NODULAR RECOGNITION METHOD BASED ON TEXTURE ANALYSIS

In this paper, a two-stage model was used to locate each suspected nodule based on known lung CT images, and the results were output to obtain a potential nodule set [7]. By selecting the candidate nodule set, the false positive nodules are eliminated. According to the 3D features of lung CT images, 3D mesh structure was used to identify candidate lesions and eliminate false positive lesions as shown in Figure 1. The image is first cut to the appropriate size, and then the possible image is extracted. After identifying the possible nodular regions, the three-dimensional dual-path network was used for identification and false positives were eliminated[8].

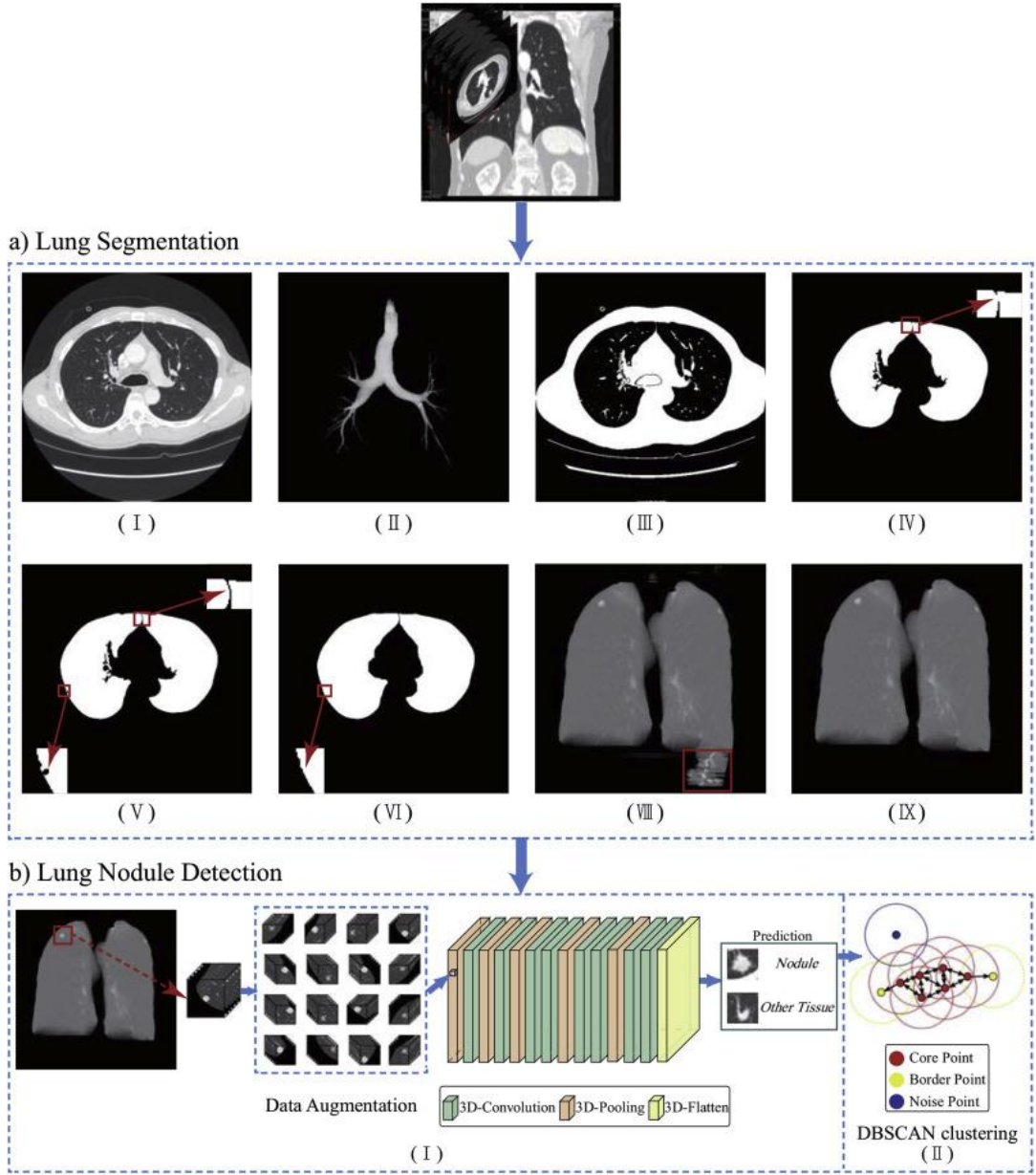

Fig. 1. Lung nodule detection model

*A. Discovery of suspicious nodules*

Faster R-CNN is a common target recognition method [9]. Like the previous fast R-CNN, it is improved based on RCNN, aiming at integrating target detection into neural networks. By deep convolution of images and selecting suitable candidate points, FastR-CNN can reduce the repeated calculation of image characteristics and speed up the learning speed. Faster-CNN uses a convolutional neural network based on region generation to replace the fast R-CNN selection, and integrates the entire training process into the architecture neural network, thus greatly improving the execution efficiency of the algorithm [10]. A multi-dimensional CNN model is proposed for image preprocessing based on the excellent performance in target identification and the spatial continuity of lung cancer CT images [11]. Taking a continuous image in CT coordinate system as an object, three-dimensional reconstruction is carried out[12]. Limited by the computing power of GPU, this project intends to use VGG-16 convolutional network for feature extraction based on 3D reconstruction of 96*96*96 pixels. Its implementation is as follows:

*1) Local creation network*

The method takes CT images as the target boxes (ROIs), and is associated with each ROI and outputs the corresponding target score [13]. The network is completely connected to an n* n spatial window [14]. The boundary regression layer and the boundary Division layer. n is 3, and the structure of the network is shown in Figure 2 (picture cited in Entropy 2023, 25(6), 837).

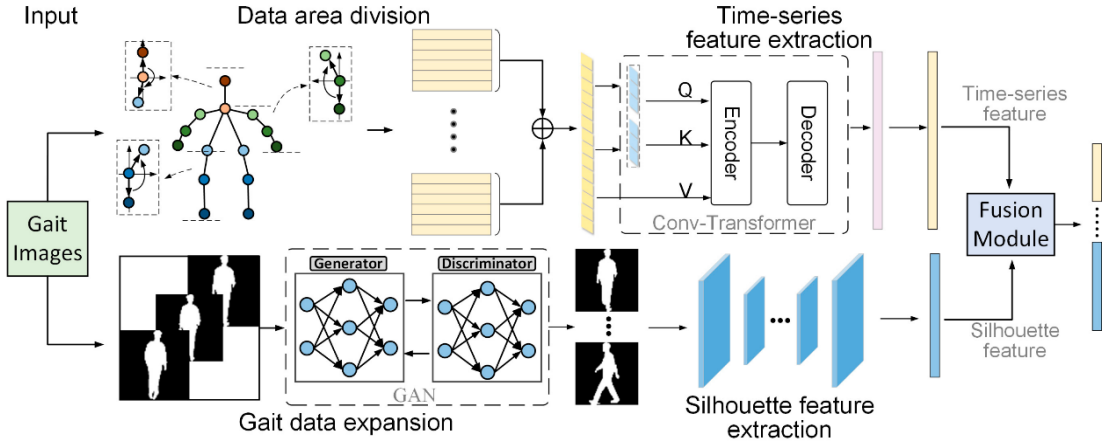

Fig. 2. Zone generation network

Because the entire small network operates as a sliding window, the entire connectivity layer shares all the spatial units [15]. This structure is essentially composed of an n* n convolutional layer and two 1*1 convolutional neural networks [16].

*2) Division of target areas*

The algorithm first uses a small convolutional neural network to extract each region of interest, then adds a large pooling layer to each region, and finally obtains the feature vector of each region of interest.

*B. Eliminate the pseudo-positive effect*

The first step is to use 3D Faster R-CNN to generate a candidate that contains a number of faulty nodes. The second step is to eliminate those false positive feedback signals [17]. In this project, two channel models based on ImageNet are used to identify lung cancer. Dual channel connectivity can benefit from residual networks[18]. The problem of gradient loss in deep neural networks can be overcome by using a short-cut connectivity technique as shown in Figure 3.

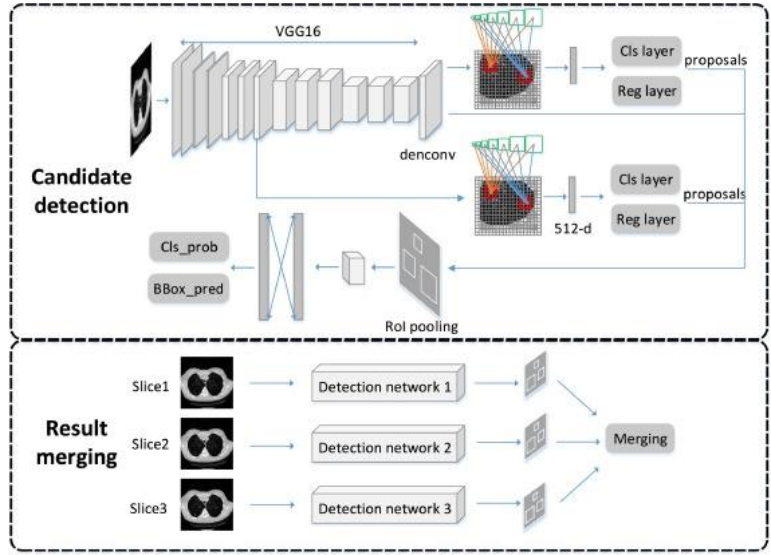

Fig. 3. Nodules identification by dual-channel network

*C. Denoise the image*

Regarding the challenges of denoising magnetic resonance imaging, the adoption of an unsupervised learning methodology can significantly aid this process by bypassing the limitations inherent in traditional supervised learning frameworks, which typically require paired training data. Utilizing an unsupervised learning approach for the denoising of magnetic resonance imaging (MRI) significantly enhances the denoising process[19].

## III. LUNG NODULE DETECTION MODEL

In the examination of pulmonary nodules, there may be an elevated false-positive. A 3D VGG network of fused residual structures was established to effectively remove false-positive nodules [20]. The VGG network is used to construct residual relation, and the maximum pool drop sampling method is used to sample each feature layer, and 4 feature layers are obtained [21]. Through the software activation function to classify the tumor, it can be divided into two types: good and bad [22].

$focal\ loss$ is introduced as the network. $focal\ loss$ is defined as

$$focalloss(g_t) = -\eta(1-g_t)^\varsigma \log(g_t) \quad (1)$$

$g_t$ is how close the classification to be predicted is to the sample. $(1-g_t)^\varsigma$ is the regulator that adjusts the weight of the difficulty sample [23]. Since only one type of pulmonary nodule was detected in this article, the average accuracy AP is an average of the accuracy of a certain type, $mAP = AP$. Set the exact ratio as a horizontal axis, the recall rate as a vertical axis, and the area below the curve as AP.

$$AP = \int_0^1 g(t)ds \quad (2)$$

This project intends to use the image information collected from patient images as samples, compare and analyze the twin sub-samples obtained by the two methods, and adopt the method based on random mean gradient to realize the accurate identification of twins [24]. The detected model parameters are used as the solution space to test each parameter [25]. If yes, the optimization result is introduced into the model; if no, the corresponding correction is carried out, and so on until the method can reach the accuracy of solving the problem.

IV. ANALYSIS OF EXPERIMENTAL RESULTS

The LUNA16 database tests the algorithm, and uses the internationally accepted FROC algorithm to evaluate it [26]. The LUNA16 library included CT images of 888 patients with a total of 1186 nodules, with 4 radiologists using artificial markers of 3-30 mm for reference [27]. Using LIDC-IDRI samples as samples, the ability of the established method to identify pulmonary nodules was verified. The LIDC-IDRI repository comprises 1018 computed tomography scans, varying in slice thickness from 0.6 mm up to 5.0 mm, with an average thickness of 2.0 mm. Figure 4 shows the various sizes of the nodal shapes.

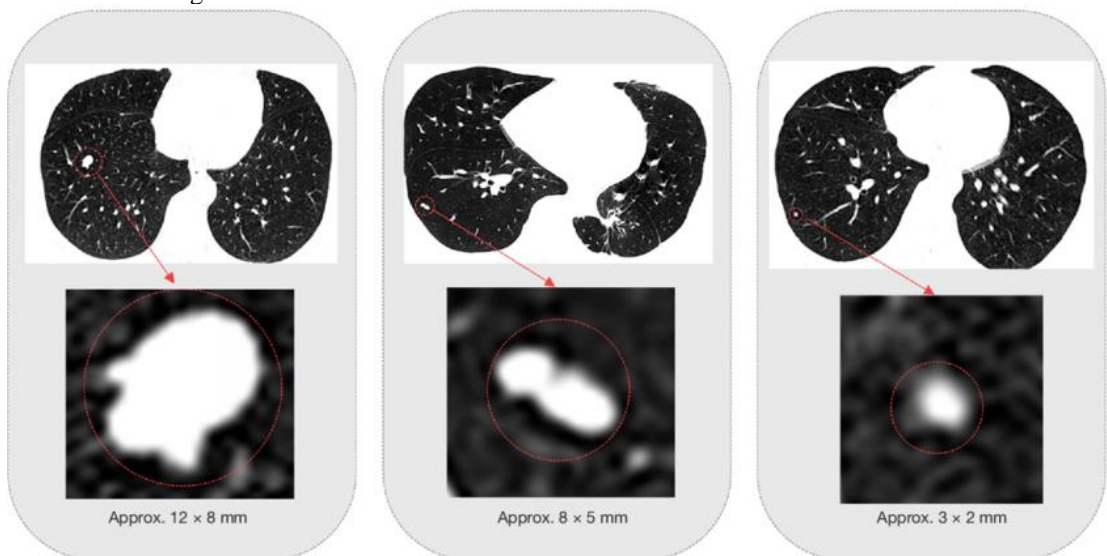

Fig. 4. Images of pulmonary nodules of different sizes

Luna test is the most representative and most popular diagnostic standard for pulmonary diseases [28]. It was tested on the Windows 10 platform with the GXT Ge Force 980 GPU. Through the positioning and localization of candidate nodules, the relative position between candidate nodules and real nodules is located to determine whether nodules exist. Table 1 shows that in the process of node discovery.

TABLE I. COMPARISON OF NODULAR DETECTION RESULTS IN EACH MODEL

| Model | Recall rate / % | Number of nodules/n |
|---|---|---|
| ISI CAD | 0.892 | 350 |
| Subsolid | 0.376 | 303 |
| Large | 0.331 | 50 |
| M5L | 0.800 | 23 |
| Ours | 0.898 | 21 |

The ISICAD model determined the shape index and camber index related to the node. Through the statistics of various parameters of CT images at all levels, the threshold screening was set to obtain suspicious "seeds", and then the adjacent "seeds" were fused to construct suspicious nodular aggregates. Large CAD algorithm is to fuse adjacent suspicious regions and finally get candidate regions. As shown in Figure 1, this method has a smaller probability of node candidate points on the premise of maintaining a high recall rate.

Figure 5 shows the FROC figures for the various models with seven different average false positives. The false positive rate refers to the number of false positives detected by CT imaging.

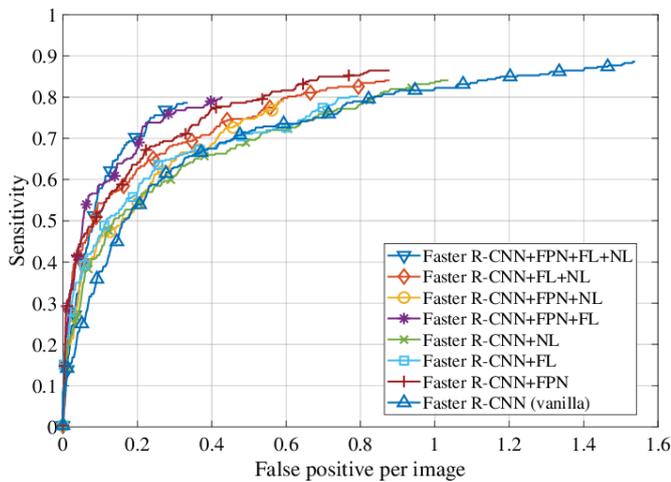

Fig. 5. FROC curves of different models

## V. CONCLUSION

This project introduced automatic lung nodule recognition technology based on the deep convolutional neural network into CT images and divided it into two stages: predicting potential lesions and eliminating false positives. This project intends to use 3D-ASTERR-CNN method to segment each nodule in 3D to obtain 32*32*32-pixel 3D image, and then use a three-dimensional dual-path network to extract its features, and use a binary Logistic regression method to identify them and eliminate false nodules. The characteristics of tumors can be better extracted based on the three-dimensional characteristics of CT images and their continuous characteristics in space. Compared with other methods based on two-dimensional CNN, this method can extract the reconstructed image information from CT images better, and the experiment is carried out on LUNA16. This method can not only help doctors in the early diagnosis of lung lesions, but also can be extended to other 3D images, so it has important significance in practical work.